\documentclass[showpacs,preprintnumbers,amsmath,amssymb]{revtex4}
\usepackage{graphicx,epsfig}

\def\Journal#1#2#3#4{{#1} {\bf #2}, #3 (#4)}

\def\PRL{Phys. Rev. Lett.}

\def\PRD{{Phys. Rev.} D}

\def\AA{{Astron. Astrophys.}}

\def\NPA{{Nucl. Phys.} A}

\def\CQG{{Class. Quant. Grav.}}

\def\be{\begin{equation}}
\def\ee{\end{equation}}
\def\bea{\begin{eqnarray}}
\def\eea{\end{eqnarray}}

\begin{document}

\title{The violation of the weak energy condition, Is it 
generic of spontaneous scalarization ?}

\date{\today}

\author{Marcelo Salgado}\email{marcelo@nuclecu.unam.mx}  
\author{Daniel Sudarsky}\email{sudarsky@nuclecu.unam.mx} 
\affiliation{Instituto de Ciencias Nucleares \\
Universidad Nacional Aut\'onoma de M\'exico\\
Apdo. Postal 70--543 M\'exico 04510 D.F., M\'exico}

\author{Ulises Nucamendi} \email{ulises@fis.cinvestav.mx}  
\affiliation{Instituto de
F\'{\i}sica y Matem\'{a}ticas, Universidad Michoacana de San
Nicol\'{a}s de Hidalgo,\\
Edif. C-3, Ciudad Universitaria, Morelia, Michoac\'{a}n, C.P.
58040, M\'{e}xico}




\begin{abstract}
It was recently shown by Whinnett \& Torres \cite{WT} that the 
phenomenon of spontaneous scalarization (SC)
in compact objects (polytropes) was accompanied also by a 
{\it spontaneous violation of the weak energy condition} (WEC). 
Notably, by the encounter of negative-energy densities at several star places as measured by 
a static observer. Here we argue that the negativeness of such densities 
is not generic of scalar tensor theories 
of gravity (STT). We support this conclusion by numerical results within 
a class of STT and by using three realistic models of dense matter. 
However, we show that the ``angular parts'' of the additional conditions 
for the WEC to hold $\rho_{\rm eff} + T^{i\,({\rm eff})}_{\,\,\,i} \geq 0$ tend to 
be ``slightly violated'' at the outskirts of the star.
\end{abstract}

\pacs{04.50.+h, 04.25.Dm, 97.60.Jd}
\maketitle

\section{Introduction}

Scalar-tensor theories of gravity (STT) are one of the simplest 
alternatives to 
Einstein's general relativity (GR) \cite{damour0}. 
A particular example of these, is the 
Brans-Dicke theory. The general feature of STT is the introduction of 
a new fundamental scalar field which couples to gravity in a non-minimal 
fashion (NMC). Nonetheless, the NMC is such that it preserves the 
Equivalence Principle, because the scalar-field couples to 
the ordinary matter only through the spacetime metric.

The NMC in STT unlike simple Einstein-Higgs systems, gives rise to 
field equations that can differ substantially from the usual 
Einstein's equations. Therefore, when applied to the astrophysical and 
cosmological settings these theories can produce severe deviations from the 
general relativity predictions. The fact that STT introduce in general only 
a few new parameters make them in principle easily testable. 
Perhaps the most notable probes in this direction are those connected 
with the binary pulsar, the primordial nucleosynthesis, 
the Cosmic Microwave Background, and the 
luminosity distance with Type-II Supernova. 
The success of GR lies in the fact that 
with suitable matter models, 
the theory is compatible with the available information in the above systems. Modifications 
of the gravitational sector in the way of STT lead thus to a 
serious change in the geometry-matter relation that can upset the 
compatibility of the theory with the observations. 
The stringent bounds on STT 
make difficult to accommodate these theories to all the observations 
while at the same time predicting new observable phenomena. One 
of the potentially observable new phenomenon is precisely the 
phenomenon of spontaneous scalarization (SC) arising in compact objects. 
As it was first shown by Damour \& Esposito-Far\`ese \cite{damour1}, STT can 
induce non-perturbative effects in neutron stars (NS) which, roughly speaking, 
consists in a phase-transition that endows a compact star 
with a new global quantity, the {\it scalar charge}. This ``charge'' is the 
analogous of the {\it magnetization} in ferromagnets at low 
temperatures and the central-energy density 
(or equivalently the baryon mass) of the object plays in turn the role of the 
temperature in spontaneous magnetization. 
Damour \& Esposito-Far\`ese \cite{damour1,damour2} 
also shown that the consequences of SC 
can be dramatic even if the boundary-conditions (asymptotic conditions) 
for the scalar-field are chosen so as to satisfy the solar-system 
bounds. In particular, they showed that the binary-pulsar dynamics 
is very sensitive to the new phenomenon, and therefore, that STT can be 
further constrained even if they successfully pass the local and the 
cosmological tests.

In this paper we will not be concerned with the confrontation 
of STT with the experiments, but rather with a result obtained by 
Whinnett \& Torres \cite{WT} who presented evidence that the 
the phenomenon of 
SC in NS is linked to a violation of the weak energy condition (WEC). 
Notably, Whinnett \& Torres \cite{WT} found that the 
effective energy density of STT as measured by a static observer and 
defined as $\rho_{\rm eff}:=n^a n^b G_{ab}/(8\pi)$ 
(where $G_{ab}$ is the Einstein tensor and $n^a$ a time-like vector parallel to the 
static Killing vector $\xi^a_t$) is negative and suggested that 
such a finding might be independent of the class of STT used (although depending 
of the chosen value of the NMC constant of the theory). 
A negative $\rho_{\rm eff}$ is a sufficient condition for the violation of the WEC.
The authors argue that the violation of the WEC in NS might led to 
their instability 
(even for masses quite below to their corresponding 
Landau-Tolman-Volkoff-Oppenheimer limit), a very undesirable
feature that would presumably rule out the STT as an alternative 
for a viable spacetime theory. 

While it is true that the effective energy density 
$\rho_{\rm eff}$ of STT as defined above is not automatically positive 
definite due to the presence of second order derivatives of the scalar field, 
the main goal of this paper is to show that there are classes of STT that do not violate the 
condition 
$\rho_{\rm eff} \geq 0$ for the problem at hand. As we shall argue in Sec. III, 
the violation of the 
condition $\rho_{\rm eff} \geq 0$ as found by Whinnett \& Torres \cite{WT} is linked to 
the use of a parametrization {\it \`a la} Brans-Dicke 
which prejudices the choice of the 
specific theory which can potentially violate such condition. In our case, we use 
a parametrization of the STT where the effective gravitational constant can be 
chosen to be positive-definite. In such a case, as it is shown in Sec. III, 
the theories parametrized in both ways may be 
very different (the transformation that would be necessary 
to change from one parametrizaction to the other 
might not be well defined). 

Altough the class of STT analyzed by us 
satisfy the 
condition $\rho_{\rm eff} \geq 0$ for the NS of interest, 
they are prone to 
violate ``slightly'' the angular parts of the additional conditions 
$\rho_{\rm eff} + T^{i\,({\rm eff})}_{\,\,\,i} \geq 0$ for the WEC to hold. 
Since this violation is 
rather weak and occurs when 
the energy-density and pressure of the star fluid is very low (near the star surface), 
we argue that the stability of the neutron star is not in jeopardy, and therefore that 
a large class of STT cannot be considered as unviable spacetime theories only by this fact.

\section{The model}
The general action for a STT with a single scalar field is given by
\be
  \label{jordan}
S = \int \left\{ \frac{1}{16\pi} F(\phi) R 
-\left( \frac{1}{2}(\nabla \phi)^2 + V(\phi) \right) \right\} \sqrt{-g} d^4x 
+ S_{\rm matt}\,\,\,.
\end{equation}
where the matter sector will be represented by a perfect fluid.

 The field equations obtained from the above action are
\bea
G_{ab} &=& 8\pi T_{ab}^{\rm eff}\,\,\,\,, \\
\label{effTmunu}
T_{ab}^{\rm eff} &=& G_{\rm eff}\left( T_{ab}^F + T_{ab}^{\phi} + T_{ab}^{\rm matt}\right)\,\,\,\,, \\
T_{ab}^F &= & \frac{1}{8\pi}\left[\nabla_a\left(\partial_\phi
F\nabla_b\phi\right) - g_{ab}\nabla_c \left(\partial_\phi
F\nabla^c \phi\right)\right] \,\,\,\,, \\
T_{ab}^{{\rm \phi}} &= & (\nabla_a \phi)(\nabla_b \phi) - g_{ab}
\left[ \frac{1}{2}(\nabla \phi)^2 + V(\phi)\right ] \,\,\,\,, \\
T_{ab}^{\rm matt} &=& (\rho+ p)u_a u_b + g_{ab}p \,\,\,,
\eea

\be
\label{KG}
{\Box \phi} = 
\frac{ F\partial_\phi V- 2(\partial_\phi F) V + 
\frac{1}{2}(\partial_\phi F)\left[ T_{\rm matt} - 
\left( 1 + \frac{3\partial^2_{\phi\phi} F}{8\pi}
\right)(\nabla \phi)^2 \right] }{ F + 
\frac{ 3(\partial_\phi F)^2}{16\pi} }\,\,\,\,.
\end{equation}
where $T_{\rm matt}$ stands for the trace of $T^{ab}_{\rm matt}$,
$G_{ab}=R_{ab} -\frac{1}{2}g_{ab}R$, and 
$G_{\rm eff} = \frac{1}{F}$. 

It is to be emphasized that the action (\ref{jordan}) corresponds to 
STT in the physical (Jordan) frame, as opposed to the one written in 
terms of an unphysical (conformally transformed) metric (the Einstein 
frame). Hereafter we consider $V(\phi)\equiv 0$. 

In a previous work \cite{us} (hereafter SSN), we analyzed the SC phenomenon in detail 
for the class of STT given by $F(\phi)= 1 + 16\pi \xi \phi^2$ 
(here $\xi$ stands for the NMC constant) 
and for spherically symmetric (static) neutron stars. In that analysis, 
we exhibited the SC phenomenon for the cases $\xi=2,6$ and employed three 
representative realistic equations of state (EOS) for the nuclear matter: 
the model of Pandharipande \cite{pand} (hereafter PandN; representative of a ``soft'' EOS), 
the model II of D\'\i az-Alonso \cite{diaz} (hereafter DiazII; 
representative of a ``medium'' EOS), and the model ``0.17'' 
of Haensel et {\it al.} \cite{hkp} (hereafter HKP; representative of a 
``stiff'' EOS). Our conclusion in SSN was that the SC is a phenomenon that ensues independently of 
the details of the EOS, but that strongly depends on the compactness of the 
star (high central energy-density). Moreover, we provided a heuristic analysis 
that clarifies why NS configurations with SC are energetically preferred 
over those with a trivial scalar field, despite the unintuitive 
expectation coming from the fact that the effective gravitational constant 
$G_{\rm eff}=1/F$ decreases as the scalar field departs from the trivial 
configuration, which 
in turn leads to a reduction of the 
absolute value of the star's negative binding energy.

In order to analyze the SC phenomenon we computed hundreds of NS configurations. 
Now, to inquire about the status of the WEC, we take representative data of our 
catalogue and recompute the energy-density profiles $\rho_{\rm eff}(r)$ 
as well as the other two relevant quantities $\rho_{\rm eff} + T^{r\,({\rm eff})}_{\,\,\,r}$ and 
$\rho_{\rm eff} + T^{\theta\,({\rm eff})}_{\,\,\,\theta}$ (where the spherical symmetry 
implies $T^{\phi\,({\rm eff})}_{\,\,\,\phi}= T^{\theta\,({\rm eff})}_{\,\,\,\theta}$). 
The WEC will be violated if any of the three above quantities is negative.

 For the static and spherically symmetric 
metric $g_{ab}={\rm diag}(-N^2,A^2,r^2,r^2\sin^2\theta)$, 
the effective energy-density 
$\rho_{\rm eff}:= n^a n^b T_{ab}^{\rm eff}$ and the quantities 
$\rho_{\rm eff} + T^{r\,({\rm eff})}_{\,\,\,r}$ and 
$\rho_{\rm eff} + T^{\theta\,({\rm eff})}_{\,\,\,\theta}$, 
respectively [where $T_{ab}^{\rm eff}$ is given by Eq. (\ref{effTmunu}), 
and $n^a$ stands for the timelike unit vector associated with a 
static observer] we have
\bea
\rho_{\rm eff} &=& 
\frac{G_{\rm eff}}{1+ 192\pi\xi^2\phi^2 G_{\rm eff}}
\left[-\frac{4\xi\phi 
(\partial_{r}\phi)(\partial_{r}\tilde N)}{\tilde N A^2}
\left(1+ 192\pi\xi^2\phi^2 G_{\rm eff}\right) 
+\frac{1}{2A^2}(\partial_{r}\phi)^2 \left(1+8\xi+ 
64\pi\xi^2\phi^2 G_{\rm eff}\right) \right. \nonumber \\
\label{Eeff}
& & \left. 
+ \rho+ 64\pi\xi^2\phi^2 G_{\rm eff}\left(2\rho+ 3p \right) 
\right] \,\,\,\,, \\
\rho_{\rm eff} + T^{r\,({\rm eff})}_{\,\,\,r} &=& 
\frac{G_{\rm eff}}{1+ 192\pi\xi^2\phi^2 G_{\rm eff}}
\left[-\frac{8\xi\phi \partial_{r}\phi}{A^2}
\left(\frac{1}{r} + \frac{\partial_{r}\tilde N}{\tilde N}\right)
\left(1+ 192\pi\xi^2\phi^2 G_{\rm eff}\right) \right. \nonumber \\
&& \left.
+\frac{1}{A^2}(\partial_{r}\phi)^2 \left(1+4\xi+ 
128\pi\xi^2\phi^2 G_{\rm eff}\right) 
+ \rho+ p + 128\pi\xi^2\phi^2 G_{\rm eff}\left(\rho+ 3p \right) 
\right] \,\,\,\,,
\label{WEC2} \\
\rho_{\rm eff} + T^{\theta\,({\rm eff})}_{\,\,\,\theta} &=& 
G_{\rm eff}\left[\frac{4\xi\phi \partial_{r}\phi}{A^2}
\left(\frac{1}{r} - \frac{\partial_{r}\tilde N}{\tilde N}\right)
+ \rho + p \right] \,\,\,\,.
\label{WEC3}
\eea
which explicitly shows the usual contributions of the 
the perfect-fluid ($\rho$ and $p$) plus  
those arising due to the NMC. Here $\tilde N=N/N_0$ is the 
renormalized lapse with respect to its value at $r=0$. 
Clearly when $\phi(r)\equiv 0$ , 
one simply recovers $\rho_{\rm eff}= \rho$, and 
$\rho_{\rm eff} + T^{r\,({\rm eff})}_{\,\,\,r}= \rho + p
= \rho_{\rm eff} + T^{\theta\,({\rm eff})}_{\,\,\,\theta}$.

\begin{figure}
\centerline{
\epsfig{figure=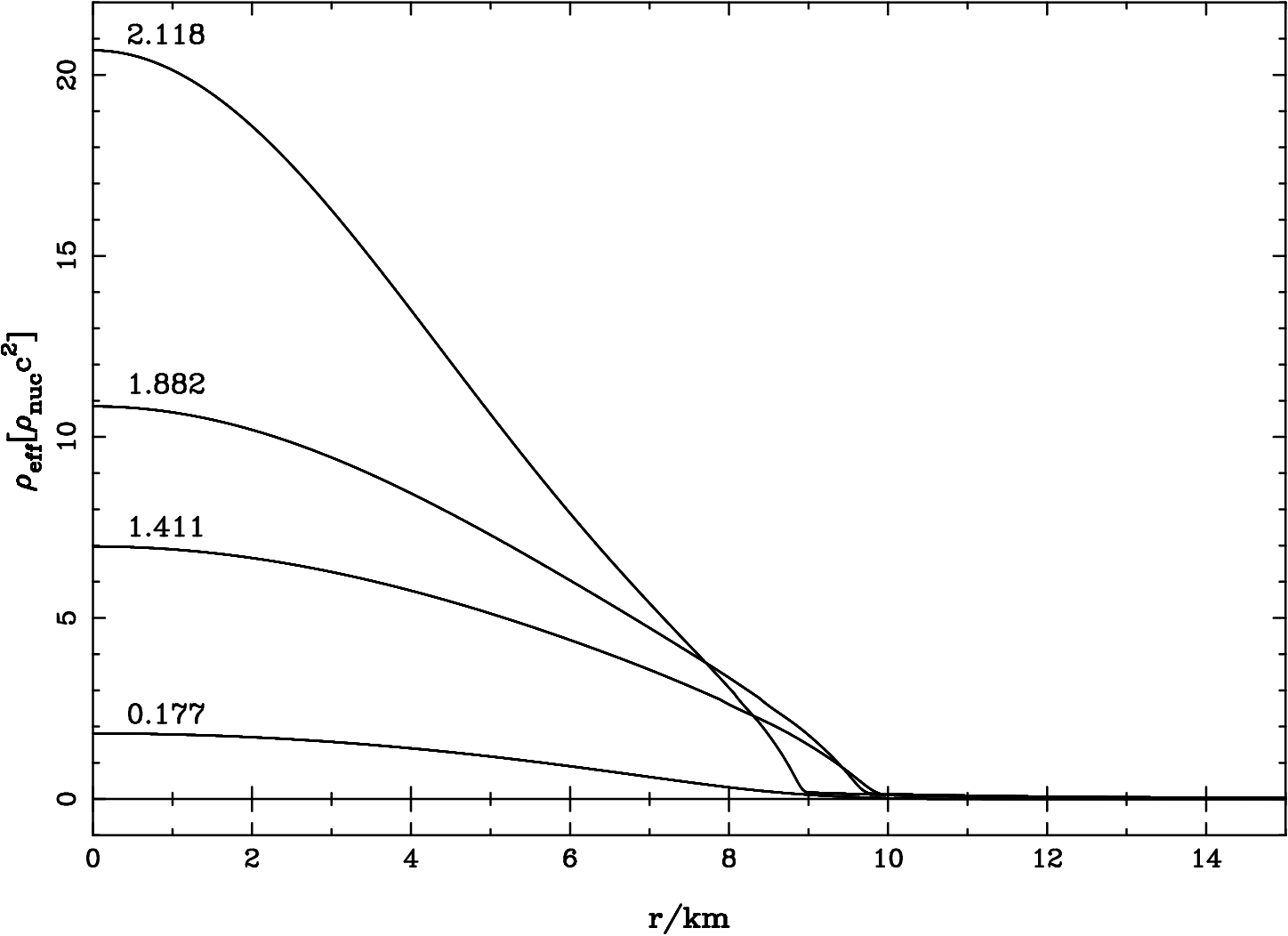,width=7cm,angle=-90}
\epsfig{figure=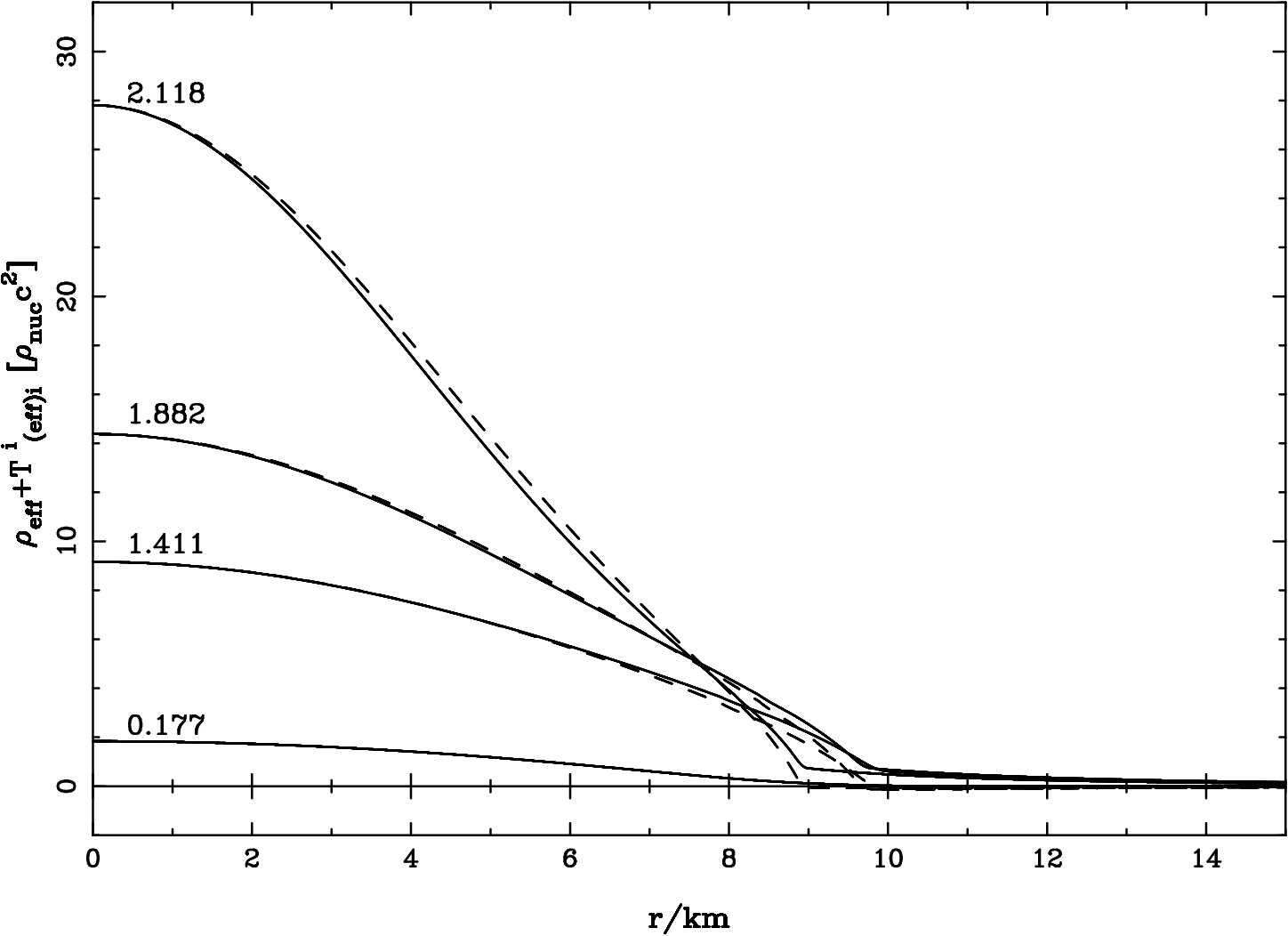,width=7cm,angle=-90}
}
\caption{}
Neutron star models constructed with the EOS PandN within the scalar-tensor theory
$F(\phi)= 1 + 16\pi \xi \phi^2$, with $\xi=6$. {\it Left panel:} 
effective energy-densities $\rho_{\rm eff}$. {\it Right panel:}
$\rho_{\rm eff} + T^{r\,({\rm eff})}_{\,\,\,r}$ (solid lines) and 
$\rho_{\rm eff} + T^{\theta\,({\rm eff})}_{\,\,\,\theta}$ (dashed lines).
 In both panels the curves are labeled 
by the corresponding baryon-mass 
(in solar mass units). The lowest baryon-mass configuration 
mark the onset of spontaneous-scalarization. The figure also shows 
the maximum mass models.
Here $\rho_{\rm nuc}= 1.66\times 10^{17} {\rm kg\, m^{-3}}$. \label{f:1}
\end{figure}

\begin{figure}
\centerline{
\epsfig{figure=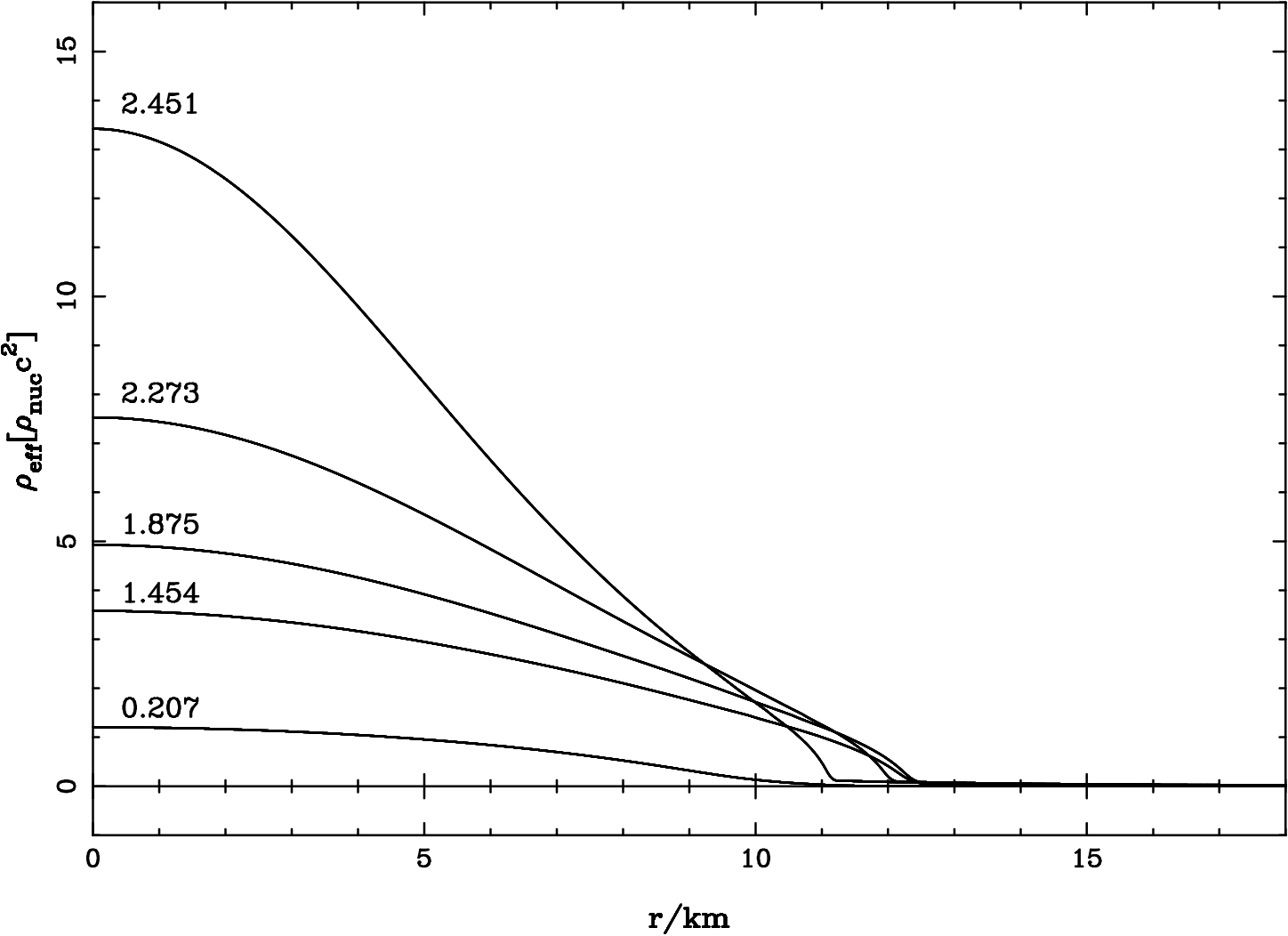,width=7cm,angle=-90}
\epsfig{figure=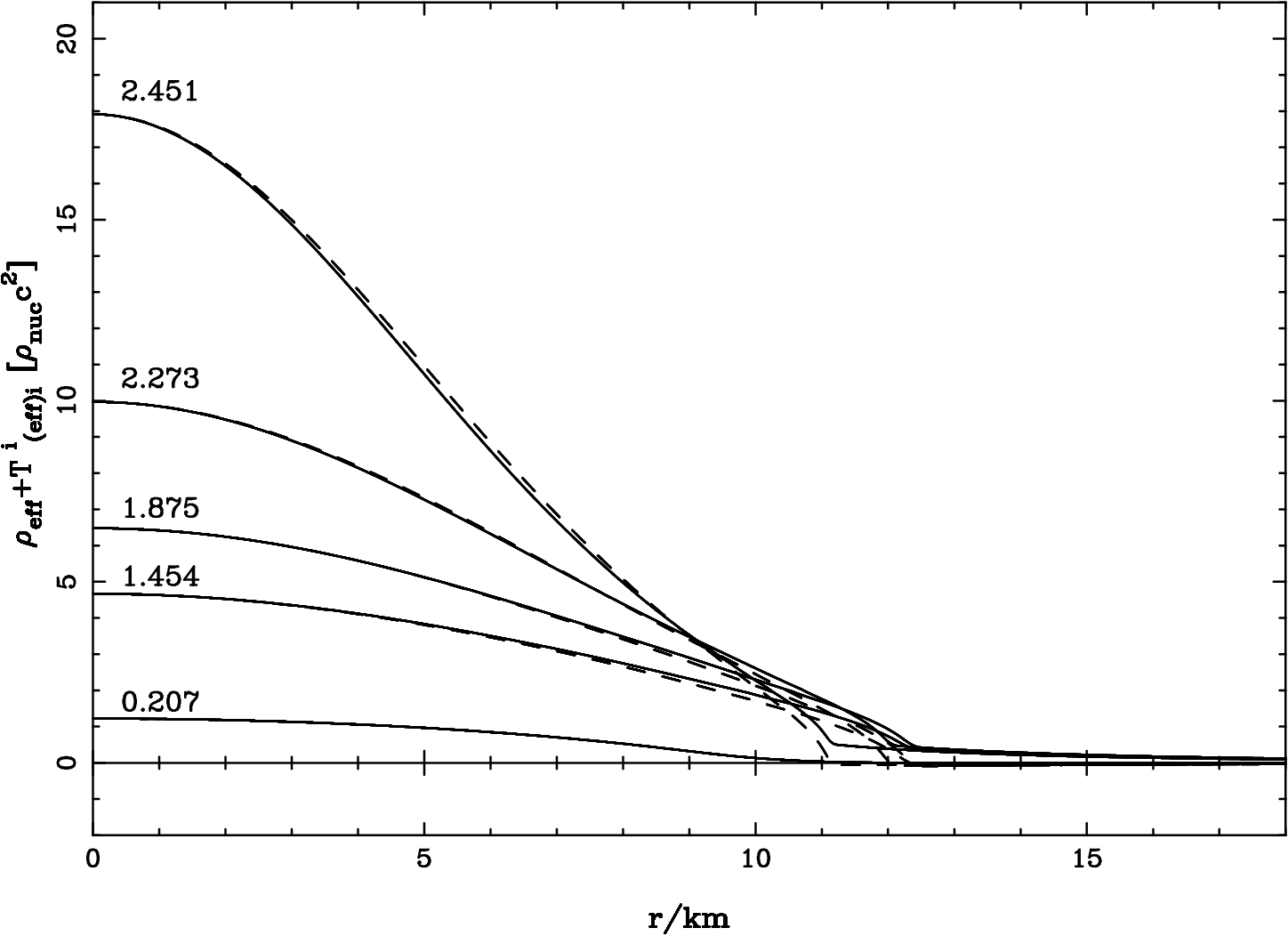,width=7cm,angle=-90}
}
\caption{}
Same as fig. \ref{f:1} for the EOS DiazII. \label{f:2}
\end{figure}

\begin{figure}
\centerline{
\epsfig{figure=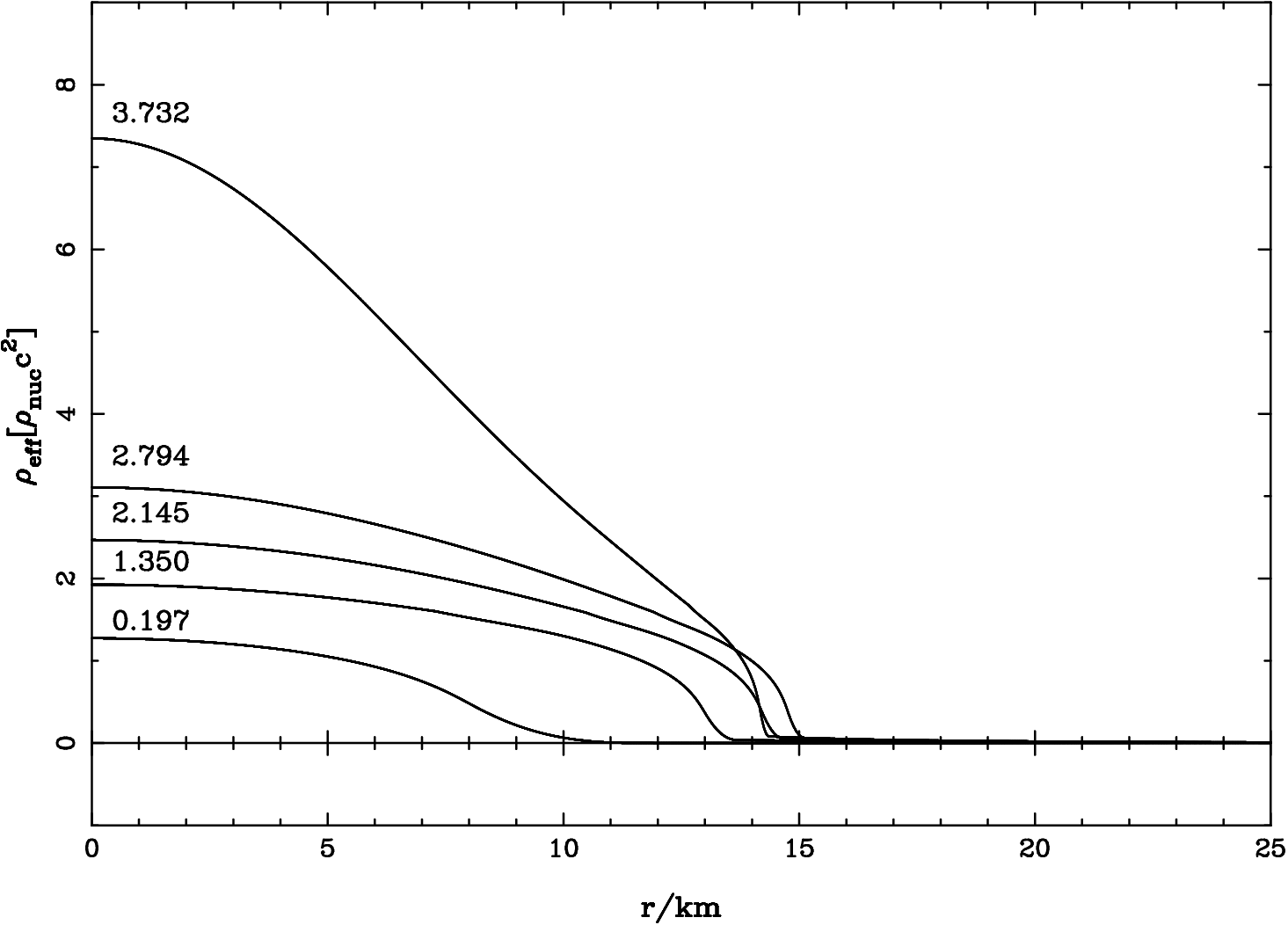,width=7cm,angle=-90}
\epsfig{figure=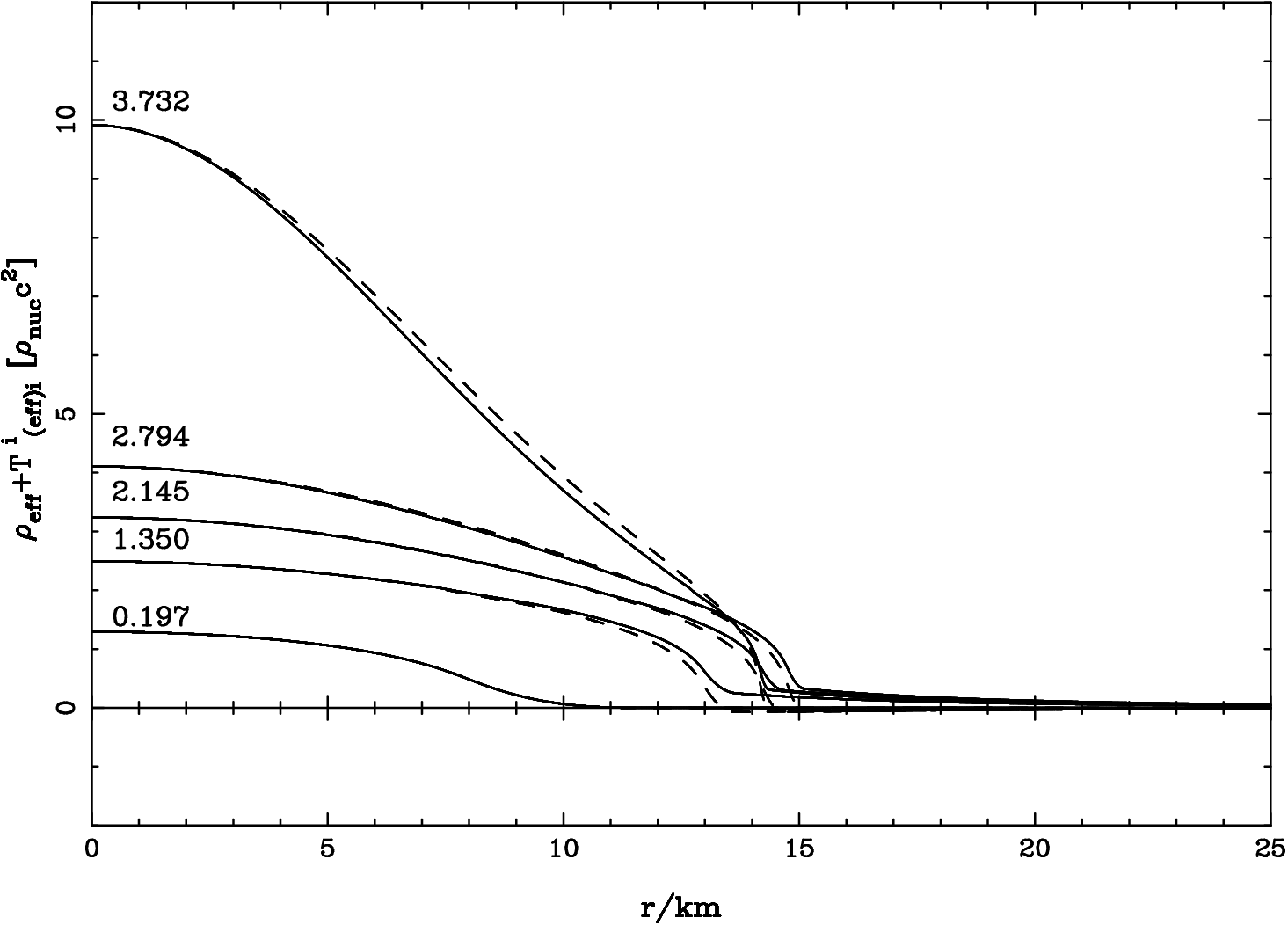,width=7cm,angle=-90}
}
\caption{}
Same as fig. \ref{f:1} for the EOS HKP. \label{f:3}
\end{figure}

We have integrated the system of equations with adequate 
regularity and boundary conditions 
(leading to asymptotically flat spacetimes) and 
with $\phi(r\rightarrow\infty)\sim Q_s/r$ where $Q_s$ is the scalar ``charge.'' 
For a fixed $\xi$, all the NS models are parametrized by the 
central energy-density of the fluid $\rho_0$. Therefore a 
critical value $\rho_0^{\rm crit}$ (or equivalently $M_{\rm bar}^{\rm crit}$)
characterizes the onset of SC. 
Figures \ref{f:1}$-$\ref{f:3} depict $\rho_{\rm eff}$ (left panels) and 
$\rho_{\rm eff} + T^{r\,({\rm eff})}_{\,\,\,r}$ (right panels; solid curves) and 
$\rho_{\rm eff} + T^{\theta\,({\rm eff})}_{\,\,\,\theta}$ (right panels; dashed curves). 
The lower mass ($M_{\rm bar}^{\rm crit}$) curves of Figs. \ref{f:1}$-$\ref{f:3} 
indicate the 
critical configurations towards SC while the rest of the curves correspond to NS endowed 
with a ``scalar charge'' $Q_s$.

The important point to note is that none of the curves of Figs. 
\ref{f:1}$-$\ref{f:3} 
show a violation of the condition $\rho_{\rm eff} \geq 0$. From the figures 
we appreciate that the condition 
$\rho_{\rm eff} + T^{r\,({\rm eff})}_{\,\,\,r} \geq 0$ is also satisfied.
However, the quantity 
$\rho_{\rm eff} + T^{\theta\,({\rm eff})}_{\,\,\,\theta}$ start becoming ``slightly'' 
negative near the surface of the star.

As we have stressed above, $\rho_{\rm eff}$ has not an a priori definite sign, 
and in fact, the term linear in $\partial_{r}\phi$ 
could be negative within the NS. Moreover, one can define 
an energy-density that is useful to understand the appearance of the phenomenon 
of SC, and is the one given as follows \cite{us},
\be
\rho^\xi =
\frac{ G_{\rm eff}}{1+ 192\pi\xi^2\phi^2  G_{\rm eff}}
\left[ \rho+ 64\pi\xi^2\phi^2 G_{\rm eff}\left(2\rho+ 3p \right)
\right] -  \rho\,\,\,\,.
\label{densroxi}
\end{equation}
which heuristically encompasses the contribution of the NMC 
to the energy-density of the perfect fluid.
 Note, however, 
that $\rho_{\rm eff}= \rho + \rho^\xi + ...$ with $\rho + \rho^\xi>0$. 
Therefore $\rho_{\rm eff} > 0$, since typically 
$|-\phi(\partial_r\phi)(\partial_r\tilde N)/(\tilde N A^2)|\ll \rho$ 
within the NS. 
In our heuristic analysis (SSN) which was also confirmed by the numerical solutions, 
we showed that the negative contribution $\rho^\xi$ more than compensates 
the decrease in the absolute value of the binding energy due to the reduction of 
$G_{\rm eff}$, 
resulting in an overall lower total energy [the ADM-energy 
that is obtained from the integration of Eq. (\ref{Eeff})] 
as compared with the energy of the configuration with the same baryon-mass 
but with a null 
scalar field (a configuration in pure GR).

\section{Discussion}
In order to give an insight for the non-negativeness of the effective energy-density 
let us consider eq. (\ref{effTmunu}) without specifying $F(\phi)$
and compute $\rho_{\rm eff}=n^a n^b T_{ab}^{\rm eff}$:
\be
\label{Eff}
 \rho_{\rm eff}= \frac{1}{F}\left[ \frac{\partial_\phi F}{8\pi}n^a n^b \nabla_a\nabla_b\phi 
+ \frac{ \frac{(\nabla\phi)^2}{2}\left(1 + \frac{(\partial_{\phi} F)^2}{16\pi F} 
+ \frac{\partial^2_{\phi\phi} F}{4\pi}\right)+ \rho + 
 \frac{(\partial_\phi F)^2}{16\pi F}\left(2\rho + 3p\right)} 
{1 + \frac{3(\partial_\phi F)^2}{16\pi F}}\right]  \,\,\,.
\end{equation}
This reduces to Eq. (\ref{Eeff}) for $F(\phi)= 1 + 16\pi \xi \phi^2$ and for 
the static and spherically symmetric case.
 
For usual nuclear matter ($\rho>0$ and $p>0$) and for $F>0$, 
$\partial^2_{\phi\phi} F> 0$, only the first 
term is not positive semi-definite. Whinnett \& Torres \cite{WT} have 
found a negative $\rho_{\rm eff}$ at various star positions, notably 
at the center of the star ($r=0$). 
So let us first focus on $\rho_{\rm eff}$ 
at $r=0$. By regularity at the origin, $\partial_r\phi|_{r=0}=0$, therefore 
 \be\label{Eff0}
\rho_{\rm eff}^0 = \frac{1}{F_0}\left[
\frac{\rho_0 + 
 \frac{(\partial_\phi F)^2_0}{16\pi F_0}\left(2\rho_0 + 3p_0\right)} 
{1 + \frac{3(\partial_\phi F)^2_0}{16\pi F_0}}\right]  \,\,\,.
\end{equation}
We can now investigate in which situations $\rho_{\rm eff}^0$ can be negative 
\footnote{It is important to note that $\phi(r)={\rm const.}\neq 0$ cannot solve the Klein-Gordon 
Eq. (\ref{KG}) within a neutron star. Therefore this excludes the 
possibility of a neutron star configuration with a
non-zero but homogeneous scalar-field. 
As a consequence, once a $\phi_0\neq0$ ensues at 
the center of the star 
(this depends on the value of the central energy-density) the only possibility 
is an inhomogeneous scalar-field which interpolates between $\phi_0$ and 
the asymptotic value. This is another way to view the onset of the spontaneous 
scalarization.}.

a) STT with $F(\phi)>0$ (in particular $F_0>0$). Clearly for this case, 
$\rho_{\rm eff}^0>0$ (since we consider only $\rho_0>0$ and $p_0>0$). 
This includes the class of STT 
we analyzed numerically $F(\phi)= 1 + 16\pi \xi \phi^2\,,\,\xi> 0$, for which we 
did not encounter negative effective central energy-densities [cf. 
Figs. \ref{f:1}$-$\ref{f:3} (left-panels)].

b) STT with $F(\phi)<0$ for some $\phi$ 
(in particular $F_0<0$). 
Clearly $F_0<0$ is a necessary condition for $\rho_{\rm eff}^0 <0$. 
One can then obtain $\rho_{\rm eff}^0<0$ if in addition the following inequalities 
hold $\rho_0 + 
 \frac{(\partial_\phi F)^2_0}{16\pi F_0}\left(2\rho_0 + 3p_0\right) >0$ 
and $1 + \frac{3(\partial_\phi F)^2_0}{16\pi F_0} > 0$. From this, we have the following 
two possibilities: 1) If $\rho_0 - 3p_0 \geq 0$ then the two inequalities hold provided 
$16\pi|F_0|> 3(\partial_\phi F)^2_0$; 
2) If $3p_0 -\rho_0 \geq 0$ then the two inequalities hold provided  
$16\pi |F_0|> C(\partial_\phi F)^2_0$ where $C= 2+ 3p_0/\rho_0$. 

According to `b)', it is in principle 
possible to obtain $\rho_{\rm eff}^0<0$ for $F_0<0$. However, 
this case is completely pathological: on one hand at the center of the star 
$F_0$ is negative (which in turn implies a negative effective gravitational 
constant $G^0_{\rm eff}=1/F_0<0$). On the other hand, asymptotically 
$F=G^{-1}_{\rm eff}\sim 1>0$ (i.e. $G_{\rm eff}$ approaches the Newtonian 
value) in order to pass successfully the solar-system 
bounds. Therefore $F(r)=F(\phi(r))$ must interpolate between the negative and the 
positive value, implying that at some spherical shell 
(inside or outside the star) $F=0$ and thus $G_{\rm eff}\rightarrow \infty$, where the equations 
become singular. 

In order to avoid this kind of pathologies within a 
neutron star model one should consider 
only the cases $F(\phi)>0$ which lead us to the conclusion that any physically viable 
STT will not produce a negative central effective-energy density
\footnote{Clearly this conclusion could dramatically change with the 
inclusion of a potential $V(\phi)$.}.

Now, although we have just proven rigorously that $F_0>0$ implies $\rho_{\rm eff}^0>0$,
we still have to proof the same for $\rho_{\rm eff}(r)$ given by Eq. (\ref{Eff}). 
Although this could be a difficult task, the 
empirical numerical evidence shows that if $\rho_{\rm eff}^0>0$ 
and for a class of STT (e.g. a class of STT with $F>0$ and $\partial^2_{\phi\phi} F> 0$) then 
$\rho_{\rm eff}(r)\geq0$ for the same class due to the fact that 
$|(\partial_\phi F)n^a n^b \nabla_a\nabla_b\phi| <<\rho$ within a NS.  

The above heuristic analysis which is supported by our numerical results leads us to 
the following conjecture:

{\bf Conjecture:} {\it Consider a STT with $F(\phi)>0$ 
and $\partial^2_{\phi\phi} F> 0$ (in particular $F_0>0$ 
at the center of a neutron star which 
implies $\rho_{\rm eff}^0>0$), 
such that asymptotically $F\sim 1$
and with $V(\phi)\equiv 0$, then 
$\rho_{\rm eff}(r)\geq 0$ for all the static NS configurations, including all those 
endowed with $Q_s$ 
(where the equality holds only outside the star and for a trivial scalar field)}
\footnote{A class of STT that can jeopardize this conjecture 
is with $F(\phi)= \chi e^{\xi \phi^2}$ ($\chi,\xi>0$), since 
$\rho_{\nabla\nabla}:= (\partial_\phi F)n^a n^b \nabla_a\nabla_b\phi\sim - 
\xi e^{\xi \phi^2} \phi (\partial_{r}\phi)(\partial_{r}\tilde N)/(\tilde N A^2)
$. So for $\xi$ large enough and with $\phi (\partial_{r}\phi)
(\partial_{r}\tilde N)> 0$ in some region within the star, 
it is possible to imagine a scenario 
with $\rho_{\nabla\nabla}<0$ and $|\rho_{\nabla\nabla}|>\rho$ so that 
$\rho_{\rm eff}$ has regions with negative values. In this case the neutron 
star would probably be unstable or even such class of STT with 
large $\xi$ would be ruled out by cosmological arguments (such as 
primordial nucleosynthesis, galaxy formation, CMB, etc.). Needless to 
say one requires further numerical studies to explore 
a concrete realization of this possibility.}.

Now, in order to give some insight about the violation of the WEC found by 
 Whinnett \& Torres \cite{WT}, let us consider their Eq. (5) (where 
$\rho_{\rm eff}$ is denoted by $\mu$) and 
focus on the central energy density 
(where the regularity conditions 
and the spherical symmetry implies that the gradient contributions become null):
\be\label{mu}
 \mu_0= \frac{1}{\Phi_0}\left[\rho_0 + \frac{(3p_0-\rho_0)}{2\omega_0(\Phi_0) +3}\right]\,\,\,.
\end{equation}
This expression is in fact equivalent to our Eq. (\ref{Eff0}), 
which we showed to be positive definite for $F_0>0$ and for usual neutron 
matter ($\rho_0>0$ and $p_0>0$). The translation 
from the Brans-Dicke parametrization used in  Eq. (\ref{mu}) to ours is provided by
\be\label{BDeqs}
\Phi = F(\phi) \,\,\,,\,\,\,
\omega(\Phi)= \frac{8\pi F}{(\partial_\phi F)^2}\,\,\,.
\end{equation}
On the other hand $\mu_0$ might not look as having 
a definite sign. We have explicitly
\be\label{mu0}
 \mu_0= \frac{1}{\Phi_0}\left[\rho_0(1-\sigma_0) + 3\sigma_0p_0\right]\,\,\,.
\end{equation}
where $\sigma(\Phi)= 1/(3+2\omega)$. Then, clearly due to the Brans-Dicke parametrization 
one could consider situations where $\sigma_0>1$ or $\sigma_0<0$, 
which depending on the values of $\rho_0$ and $p_0$ can lead 
to $\mu_0<0$. However, we note 
that the condition $F(\phi)>0$ (in particular $F_0>0$) leading to 
$\rho_{\rm eff}^0>0$ (or equivalently $\mu_0>0$), 
implies $\Phi>0$ and also $0\leq \sigma< 1/3$: In fact, 
from $\sigma(\Phi)= 1/(3+2\omega)$ and Eqs. (\ref{BDeqs}) one obtains, 
$\sigma(\Phi)= \frac{(\partial_\phi F)^2}{3(\partial_\phi F)^2 + 16\pi F}$ 
(thus $F(\phi)>0$ implies $0\leq \sigma< 1/3$). Moreover, since $\Phi= F(\phi)$ 
(assuming $F(\phi)\neq {\rm const.}$) the inverse mapping is
 $\phi= \int\sqrt{\frac{1-3\sigma(\Phi)}{16\pi\Phi \sigma(\Phi)}}\,d\Phi$. Therefore the 
conditions $\Phi>0$ and $0<\sigma< 1/3$ are required for this mapping 
$\Phi\rightarrow \phi$ to be well defined (the separate case $\sigma \equiv 0$ corresponds to 
$F(\phi)={\rm const.}$ from which one simply recovers GR). 

Hence, the restrictions on 
$\sigma$ 
bounds in turn the choice of the Brans-Dicke function $\omega(\Phi)$. 
Any choice of $\omega(\Phi)$ leading to a violation of the condition 
$0\leq \sigma < 1/3$ with $\Phi>0$ will imply a lost of the connection 
with the original theory. It is in this sense that the use of the Brans-Dicke parametrization 
prejudices the kind of STT (those STT which allows values for $\sigma$ to be outside the bounds $0\leq \sigma < 1/3$) that 
can lead to violations of the positiveness of $\rho_{\rm eff}$.

One could then use the 
SC phenomenon in NS to classify the STT in two types: Type-I) STT with $\sigma(\Phi)$ such 
that $0\leq \sigma< 1/3$ and $\Phi>0$ (which in particular are valid at the star's center 
and therefore that coincide with the case ``a)'' analyzed above). Type-II) 
STT with $\sigma(\Phi)$ (in particular $\sigma_0$)
not necessarily 
limited in the range $[0,1/3)$ and with $\Phi_0>0$.
 The first kind of STT 
can be put in correspondence with the original parametrization  
($F(\phi)>0$) and these theories lead to $0< \rho_{\rm eff}^0\equiv \mu_0>0$, 
regardless of the EOS for nuclear matter (assuming $\rho$ and $p$ non-negative). 
The type-II is a STT which cannot be put in correspondence 
with the original parametrization (in terms of $\phi$) since the 
transformation as shown above is not well defined. 
The type-II theories can lead to situations where 
$\mu_0<0$ but are not even related to the theories 
which in the original parametrization give rise to 
$\rho_{\rm eff}^0 <0$ (case ``b)'' above) which require $F_0<0$. 
For instance, 
it is possible to have $\mu_0<0$ with $\Phi_0>0$ 
with a suitable $\sigma_0$, as in the following situations:
 A) $\rho_0/(\rho_0-3p_0)<\sigma_0$ with 
$3p_0 < \rho_0$; B) $\sigma_0<0$ with 
$\rho_0/(3p_0-\rho_0)<|\sigma_0|$ and $\rho_0<3p_0$. 
Whinnett \& Torres \cite{WT} considered two classes of STT. The first one with 
$\sigma= 2\kappa {\rm ln}\Phi$ and $\kappa\geq 3$. This example can belong to Type-IIA
(and produce $\mu_0<0$) if for instance $\sigma_0 \approx 2$, $p_0<\rho_0/6$ and 
$1<\Phi_0\lesssim 1.4$. 
These conditions could be met in neutron stars with a low baryon mass 
like some of the ones computed by Whinnett \& Torres \cite{WT}.

The second class of STT used by those authors is one with 
$\sigma=\lambda(\Phi-1)$ and $\lambda \geq 5.35$. 
This class is of Type-IIA or Type-IIB
 for $\Phi_0> 1$ or 
$0<\Phi_0 <1$, respectively. As mentioned above, the condition $3p_0 < \rho_0$ 
in Type-IIA is met usually in 
NS with a low baryon mass. Since for this second class of STT, $\mu_0<0$ was found by 
Whinnett \& Torres \cite{WT} in 
low baryon mass configurations, presumably $\Phi_0> 1$ there.

We have thus given a heuristic explanation for a negative $\mu_0$ 
in both of their classes of STT.

Of course, STT of Type-II can lead to 
positive $\mu_0$'s 
without implying that the WEC will not be violated somewhere within the star (see figures of Whinnett \& Torres \cite{WT}). On the other hand, the fact that STT's of Type-I lead 
to positive definite 
$\mu_0$ does not imply that the WEC is not violated either. However, our numerical 
analysis shows that this type of STT (theories with
$\mu_0>0$) indeed satisfy the condition $\mu(r)\geq 0$ 
(where the equality holds only outside the star and for a trivial scalar field)
 and this is the motivation for our conjecture above. 
Along the same line of reasoning one could further conjecture (conjecture 2) that 
{\it given 
a fixed EOS, if a STT of Type-II violates the condition $\mu(r)\geq 0$ somewhere within a star 
(not necessarily at the center) 
then the STT will violate the WEC at the center for some 
other $\rho_0$ (defined as $\rho_0^-$)}. 
That is, the second conjecture states that $\mu_0=f(\rho_0)$ has always a negative 
branch ($f(\rho_0^-)<0$) in the Type-II theories. 
The first conjecture on the other hand roughly states 
that a STT (with the given assumptions) that 
does not violate $\mu_0>0$ (Type-I with $\mu_0(\rho_0)>0$) 
will not violate it elsewhere within the NS (i.e., $\mu(r)\geq 0$ within 
the star). 

We can now turn the attention to the condition $\rho_{\rm eff} + T^{\theta\,({\rm eff})}_{\,\,\,\theta}$ which 
the numerical analysis shows to be negative near the star surface. From eq. (\ref{WEC3}) we appreciate that 
beyond the compact support of the nuclear fluid, 
$\rho_{\rm eff} + T^{\theta\,({\rm eff})}_{\,\,\,\theta}\sim 
G_{\rm eff}\left[\frac{4\xi\phi \partial_{r}\phi}{A^2}
\left(\frac{1}{r} - \frac{\partial_{r}\tilde N}{\tilde N}\right) \right]$. If we roughly approximate 
$A^{-1}\approx N\approx (1-2M/r)^{1/2}$ for $r\geq R$ (where $R$ and $M$ stand for the star radius and total mass)
then, the factor $\left(\frac{1}{r}-\frac{\partial_{r}\tilde N}{\tilde N}\right)
\approx 1/r\left[(1 - 3M/r)/(1 - 2M/r)\right]$ which is positive even for the maximum mass configurations. 
Therefore $\rho_{\rm eff} + T^{\theta\,({\rm eff})}_{\,\,\,\theta}\sim 
G_{\rm eff}\frac{4\xi\phi \partial_{r}\phi(1 - 3M/r)}{r}$. 
Since the SC configurations have $\partial_{r}\phi<0$ near and beyond $R$
[$\phi(r)\sim Q_s/r$ asymptotically, with $Q_s>0$], 
then  $\rho_{\rm eff} + T^{\theta\,({\rm eff})}_{\,\,\,\theta}$ 
is negative (for $\xi >0$) near and beyond the star surface. Note from Eq. (\ref{WEC2}), that 
the term $\rho_{\rm eff} + T^{r\,({\rm eff})}_{\,\,\,r}$ [unlike 
$\rho_{\rm eff} + 
T^{\theta\,({\rm eff})}_{\,\,\,\theta}$; cf. Eq. (\ref{WEC3})] does not 
become negative because of the sign 
difference in the term with $\xi\phi \partial_{r}\phi$, 
and the remaining terms involving the scalar field 
are positive definite (for $\xi >0$).

In summary, there exists classes of STT that 
give rise to the phenomenon of spontaneous scalarization with and without a violation of the 
condition $\rho_{\rm eff} \geq 0$. Examples of STT like the ones we analyzed numerically does 
not violate that 
condition. However they do violate the WEC ``slightly'' 
because $\rho_{\rm eff} + T^{\theta\,({\rm eff})}_{\,\,\,\theta}$ 
starts becoming ``slightly'' negative from the star surface. 
On the other hand, there are STT like the ones analyzed by 
Whinnett \& Torres \cite{WT} which 
do violate the condition $\rho_{\rm eff} \geq 0$ which straightforwardly leads to a violation of the WEC. In our case it is unclear whether the small violation of the WEC by the STT considered 
here will necessarily lead to the instability of the NS. 
In Ref. \cite{us}, we have shown that the energetically preferred NS configurations 
(with a fixed total baryon mass) are those with lower ADM mass which correspond to scalarized NS as opposed to the corresponding 
NS in pure GR. The curves of the ADM as a 
function of the central energy-density $\rho_0$ suggest that configurations with ADM mass smaller than the maximum mass models
(corresponding to $\rho_0^{\rm max}$) 
are stable with respect to radial linear perturbations \cite{us}. 
Moreover, numerical non-linear analysis strongly suggest that ordinary NS 
(with weak or zero scalar charge) can 
decay to their corresponding more stable strong 
scalarized configurations \cite{novak2} and furthermore that among these scalarized NS
 only the configurations with 
$\rho_0>\rho_0^{\rm max}$ are unstable 
and can collapse into a Schwarszschild black hole like ordinary NS, by radiating away all the scalar-field 
\cite{novak1}.

If found that only the scalarized NS which
violate the WEC through the violation of $\rho_{\rm eff} \geq 0$ within the star
 are unstable, then the STT 
allowing such violations would be ruled out. On the other hand, it is well possible that small violations of the 
WEC through the term $\rho_{\rm eff} + T^{\theta({\rm eff})}_{\,\,\,\theta}$ in the outskirts of the NS 
while respecting $\rho_{\rm eff} \geq 0$ within the star might not lead to instabilities and therefore 
that the employed STT is a viable candidate for a spacetime theory.
 
\section{Conclusion}
Our main conclusion is that the large 
violations of the WEC deep inside the neutron stars as an additional feature to 
the phenomenon of spontaneous scalarization is not generic of all 
the classes of scalar tensor theories of gravity 
as the work of Torres \& Whinnett work might be thought to suggest. 
We have argued that there are classes of STT (at least a subset with $F(\phi)>0$ 
and $\partial^2_{\phi\phi} F> 0$) where $\rho_{\rm eff} \geq 0$ and where the 
violations of the WEC in neutron stars occur only near and beyond the 
surface of the star via the angular parts of $\rho_{\rm eff} + T^{i\,({\rm eff})}_{\,\,\,i}$. Strong 
numerical evidence supports this conclusion.

\medskip
\section*{Acknowledgments}
\medskip

M.S. and D.S. were partially supported by 
DGAPA-UNAM grants {\tt IN112401}, {\tt IN122002} and 
{\tt IN108103}. U.N. acknowledges partial support from 
SNI, and grants 4.8 CIC-UMSNH, PROMEP PTC-61, CONACYT 42949-F.

\newpage

\end{document}